\documentclass[aps,pre,superscriptaddress,noshowpacs,longbibliography,twocolumn]{revtex4-2}
\usepackage{amsmath}
\usepackage{amssymb}
\usepackage{tikz}
\usepackage{graphicx}
\usepackage{dcolumn}
\usepackage{bm}
\usepackage{epsfig}
\usepackage{psfrag}

\newcommand{\reffig}[1]{\mbox{Fig.~\ref{#1}}}
\newcommand{\refeq}[1]{\mbox{Eq.~(\ref{#1})}}
\newcommand{\refsec}[1]{\mbox{Sec.~\ref{#1}}}

\newcommand{\be}{\begin{equation}}
\newcommand{\ee}{\end{equation}}
\newcommand{\bal}{\begin{align}}
\newcommand{\eal}{\end{align}}
\newcommand{\ba}{\begin{eqnarray}}
\newcommand{\ea}{\end{eqnarray}}
\renewcommand{\Re}{\mathrm{Re}}
\renewcommand{\Im}{\mathrm{Im}}

\begin{document}

\title{Failure of the conformal-map method for relativistic quantum billiards}
\author{Barbara Dietz}
\email{bdietzp@gmail.com}
\address{Center for Theoretical Physics of Complex Systems, Institute for Basic Science (IBS), Daejeon 34126, Korea}
\address{Basic Science Program, Korea University of Science and Technology (UST), Daejeon 34113, Korea}

\date{\today}

\begin{abstract}
	In [H. Xu {\it et.al.}, Phys. Rev. Lett. {\bf 110}, 064102 (2013)] a numerical method is introduced -- an extension of the conformal-map method of [M. Robnik, J. Phys. A: Math. Gen. 17, 1049 (1984)] for nonrelativistic quantum billiards -- for the quantization of relativistic neutrino billiards (NBs) consisting of a massless non-interacting spin-1/2 particle confined to a two-dimensional domain. We demonstrate in this work, that this method does not provide solutions of the associated Weyl (Dirac) equation, nor does it fulfill the boundary conditions imposed on the spinor eigenfunctions to ensure confinement of the particle to the domain of the billiard. We review in detail the wave equation, boundary conditions and quantization of NBs and derivation of relevant equations, to make the proof comprehensible for the general reader. Our results are corroborated with numerical ones for non-relativistic and relativistic quantum billiards whose shapes depend on a parameter, which allows the study of the properties of their eigenstates as the classical dynamics experiences a transition from regular to chaotic dynamics.   
\end{abstract}
\bigskip
\maketitle

\section{Introduction\label{Intro}}
Billiards have served since the 1980th as a paradigm model for theoretical, numerical and experimental investigations of aspects of quantum chaos~\cite{Sinai1970,Bunimovich1979,Berry1981}. In the classical limit they refer to a bounded two-dimensional domain, in which a point particle moves freely and is reflected specularly at the boundary and have the particular property that their dynamics can be controlled by their shape. Nonrelativistic quantum billiards (QBs) are governed by the Schr\"odinger equation for a free particle which is confined to the billiard domain by imposing on the wave functions the Dirichlet boundary condition (BC)~\cite{LesHouches1989,Stoeckmann1990,Richter1999,Haake2018}, and can be realized experimentally with flat microwave cavities~\cite{Stoeckmann1990,Sridhar1991,Graef1992,Stein1992,So1995,Deus1995,Dietz2015a}. In 1987 Berry and Mondragon proposed relativistic neutrino billiards~\cite{Berry1987} (NBs). These are governed by the Weyl (Dirac) equation for a spin-1/2 particle which is confined to the billiard area by imposing the BC that there is no outward current. 

For the study of spectral properties of QBs and NBs generally complete sequences of several 1000th of eigenvalues are needed. Their determination can be intricate and may require sophisticated methods, especially when the classical dynamics is close to integrable implying that the spacings between adjacent eigenvalues may become much smaller than their mean value. Such methods have been developed for nonrelativistic billiards based on the finite-element method~\cite{Heuveline2003}, which in the mean time has been elaborated considerably, and for high-lying eigenstates one involving the boundary norm~\cite{Vergini1995}. The boundary-integral method (BIM) has been introduced in Ref.~\cite{Baecker2003} for QBs, in~\cite{Berry1987} for massless NBs and for massive ones in~\cite{Dietz2020,Dietz2022}. It originates from the Green theorem, which provides an exact integral equation for the eigenfunctions in the interior of a QB or NB in terms of those on the boundary, that is, the associated BCs are incorporated. In principle it is applicable to arbitrary shapes and has been applied even to dielectric billiards~\cite{Wiersig2003}, for which one has to deal with singularities~\cite{Ketzmerick2025,Yi2025}. However, modifications might be needed, e.g., when there are inner corners~\cite{Okada2005}. In the presence of nearly degenerate eigenvalues one might have to struggle in both cases with missing levels. For that case an expanded BIM has been developed in Refs.~\cite{Veble2007,Yupei2019}. Yet, for NBs, the occurrence of nearly degenerate eigenvalues is considerably less an issue than for nonrelativistic ones, and can generally be removed when they originate from discrete rotational symmetries by separating into the associated symmetry classes~\cite{Baecker2003,Dietz2023}. 

In~\cite{Robnik1984} a  conformal-map method (CMM) has been introduced for the quantization of QBs whose shapes are generated from a conformal mapping of the circle, which provides the solutions of the associated Schr\"odinger equation with high accuracy. The method uses an expansion in terms of the eigenstates of the corresponding circle QB, which complies with the Dirichlet BC along the boundary of the QB by construction. This CMM has been applied to NBs, e.g., in~\cite{Xu2013,Huang2018,Song2019,Li2022} to analyze relativistic quantum scars. We will demonstrate in this work, that the CMM does not provide the eigenstates of NBs, although the eigenvalues might be close to the exact ones for sufficiently small deformations of the circular NB.      

\section{Review of neutrino billiards\label{Review}} 
We consider two-dimensional neutrino billiards (NBs), that were introduced in Ref.~\cite{Berry1987}. A NB consists of a massless relativistic spin-1/2 particle moving in a bounded two-dimensional region and is described by the Weyl equation~\cite{Weyl1929} for Weyl fermions, which is generally referred to as Dirac equation,
\be
\boldsymbol{\hat H}_D\boldsymbol{\psi}=c\boldsymbol{\hat\sigma}\cdot\boldsymbol{\hat p}\boldsymbol{\psi}
=E\boldsymbol{\psi},\, \boldsymbol{\psi}=
\begin{pmatrix}
\psi_1 \\ \psi_2
\end{pmatrix}.
\label{DE}
\ee
Here, $\boldsymbol{\hat p}=-i\hbar\boldsymbol{\nabla}$ is the momentum of the particle, $\boldsymbol{\hat H}_D$ denotes the Dirac Hamilton operator, $\boldsymbol{\hat\sigma}=({\hat\sigma}_x,{\hat\sigma}_y)$, and ${\hat\sigma}_{x,y,z}$ are the Pauli matrices. The energy of the particle is $E=\hbar ck$, where $k$ is the free-space wave vector and $c$ is the velocity of light in vacuum. The particle is confined to the billiard domain by imposing the BC that the outward flux vanishes. Accordingly, the normal component of the local current, that is, of the expectation value of the current operator $\boldsymbol{\hat u}=\boldsymbol{\nabla}_{\boldsymbol{p}}\boldsymbol{\hat H}_D=c\boldsymbol{\hat\sigma}$ is required to vanish along the boundary~\footnote{This choice of BCs is not the only one guaranteeing self-adjointness of the Dirac Hamiltonian and zero outgoing current~\cite{Gaddah2018,Greiner1994}.},
\be
\label{Flux}
\boldsymbol{n}\cdot\left[\boldsymbol{\psi}^\dagger\boldsymbol{\nabla}_{\boldsymbol{p}}\boldsymbol{\hat H}_{D}\boldsymbol{\psi}\right]=c\boldsymbol{n}\cdot\left[\boldsymbol{\psi}^\dagger\boldsymbol{\hat\sigma}\boldsymbol{\psi}\right]=0.
\ee

We consider NBs whose domain $\Omega$ can be defined in a cartesian coordinate system in polar coordinates, $\boldsymbol{r}=[x(r,\varphi),y(r,\varphi)]$, or in the complex plane, 
\be
w(r,\varphi)=x(r,\varphi)+iy(r,\varphi), \varphi\in [0,2\pi),\, r\in [0,r_0]
\label{coordinate}
\ee
with the boundary $\partial\Omega$ at $r=r_0$ denoted as $w(\varphi)\equiv w(r=r_0,\varphi)$. Then, the BC,~\refeq{Flux}, reads~\cite{Berry1987}
\be
\psi_2(\varphi)=i\mu B(\varphi)e^{i\alpha(\varphi)}\psi_1(\varphi).
\label{BC1}
\ee
Here, $\alpha(\varphi)$ is the angle of the outward-pointing normal vector $\boldsymbol{n}=[\cos\alpha(\varphi),\sin\alpha(\varphi)]$ at $w(\varphi)$ with respect to the $x$ axis, and $\mu=\pm 1$ determines the rotational direction of the current at the boundary. We use $\mu =1$ in the following and $B(\varphi)=1$ in accordance with Ref.~\cite{Berry1987}.

The normal vector $\boldsymbol{n}$ can be expressed in terms of the derivative ${d}{w(\varphi)}/{d}\varphi=w^\prime(\varphi)$, $e^{i\alpha(\varphi)}=-i\frac{w^\prime (\varphi)}{\vert w^\prime (\varphi)\vert}$, yielding the BC
\be
\psi_2(\varphi)=\frac{w^\prime (\varphi)}{\vert w^\prime (\varphi)\vert}\psi_1(\varphi)\, .
\label{BC2}
\ee
Here, we used that for fixed $r=r_0$ we have $w^\prime(z)=-\frac{i}{r_0e^{i\varphi}}w^\prime(\varphi)$. Note, that $\frac{{d}\alpha(\varphi)}{{d}\varphi}=\kappa(\varphi)\vert w^\prime(\varphi)\vert$, with $\kappa (\varphi)$ denoting the curvature of the boundary at $\varphi$. 

We restrict throughout the work to NBs whose shapes are generated by a conformal mapping of the circle defined in terms of polar coordinates $0\leq r\leq r_0$ and $0\leq\varphi\leq 2\pi$ by a polynomial in $z$,
\be
w(z)=x(r,\varphi)+iy(r,\varphi)=\sum_{l\geq 0} c_lz^l,\, z=re^{i\varphi}, 
\label{WZ}
\ee
for $r\in\left[0,r_0\right],\, \varphi\in[0,2\pi),\, w^\prime(z)\ne 0$.
Here, the $c_l$ are real or complex coefficients. For $r<r_0$ and $r=r_0$, $w(z)$ parameterizes the billiard domain $\Omega$ and boundary $\partial\Omega$, respectively. The transformation~\refeq{coordinate} from $(x,y)$ to $w(z=re^{i\varphi})$ yields for the gradient in the complex plane
\be
\label{Grad}
\frac{\partial}{\partial x} +i\frac{\partial}{\partial y}=
\frac{1}{\left[zw^\prime (z)\right]^\ast}\left( r\frac{\partial}{\partial r} +i\frac{\partial}{\partial \varphi}\right),
\ee
where the star $^\ast$ denotes complex conjugation, leading to the Dirac equation, which consists of coupled equations for the spinor components $\psi_1(r,\varphi)$ and $\psi_2(r,\varphi)$, 
\ba
\label{Components}
\label{DGLpsi1}
ik\psi_1(r,\varphi)&&=\frac{1}{\left[z w^\prime (z)\right]}\left( r\frac{\partial}{\partial r} -i\frac{\partial}{\partial \varphi}\right)\psi_2(r,\varphi),\\
\label{DGLpsi2}
ik\psi_2(r,\varphi)&&=\frac{1}{\left[z w^\prime (z)\right]^\ast}\left( r\frac{\partial}{\partial r} +i\frac{\partial}{\partial \varphi}\right)\psi_1(r,\varphi).
\ea
Applying the differential operator given in~\refeq{Grad} to~\refeq{DGLpsi1} and its complex conjugate to~\refeq{DGLpsi2} and using that 
\ba
\left( r\frac{\partial}{\partial r} +i\frac{\partial}{\partial \varphi}\right)z=0,
\label{Derivz}
\ea
and~\refeq{Grad} gives for each spinor component a separate Schr\"odinger equation,
\be
\label{Schr}
\Delta_{(r,\varphi)}\psi_j(r,\varphi)=-k^2\left\vert w^\prime (z)\right\vert^2\psi_j(r,\varphi), j=1,2
\ee
with $\Delta_{(r,\varphi)}$ denoting the Laplace operator in polar coordinates.
Yet, when solving these wave equations, one needs to take into account that the components are linked through~\refeq{DGLpsi1} or, equivalently, through~\refeq{DGLpsi2}, and along the boundary in addition through the BC~\refeq{BC2}. 

Solutions of the Dirac equation~(\ref{DGLpsi1}) and~(\ref{DGLpsi2}) can be written in the coordinate system~\refeq{coordinate} in terms of a plane-wave expansion of the form~\cite{Dietz1993},
\be
\Phi_{1}(r,\varphi) = \sum_la_{l}(k)i^lJ_l(k\vert w(z)\vert)e^{il\theta(z)},
\label{Phi1}
\ee
where $J_l(x)$ denotes the Bessel function of the first kind, the coefficients $a_{l}(k)$ are complex or real and independent of $(r,\varphi)$ and $e^{i\theta(z)}=\frac{w(z)}{\vert w(z)\vert}$. Namely, inserting this ansatz into Eqs.~(\ref{DGLpsi1}) and~(\ref{DGLpsi2}) yields with equations~(S1),~(S2) and~(S3) in~\refsec{suppl} for the second component
\be
\label{Phi2}
\Phi_{2}(r,\varphi)=\sum_la_{l}(k)i^{l+1}J_{l+1}(k\vert w(z)\vert)e^{i(l+1)\theta(z)},
\ee
and 
\begin{align}
\label{Psic1}
&\Delta_{(r,\varphi)}\Phi_{j}(r,\varphi)\equiv\\
&-k^2\vert w^\prime(z)\vert^2\sum_la_{l}(k)i^{l-1+j}J_{l-1+j}(k\vert w(z)\vert)e^{i(l-1+j)\theta(z)},
\nonumber
\end{align}
with $j=1,2$, implying that both spinor components solve the Schr\"odinger equation for a free particle~\refeq{Schr}. To be an eigenstate of the NB with shape $w(z)$, $\Phi_{1,2}(r,\varphi)$ should fulfill at discrete values of $k$ the BC~\refeq{BC2}. Finding such solutions can be a cumbersome task. A very efficient method, which is based on the BIM, is introduced in~\cite{Berry1987}. 

In~\cite{Robnik1984} a quantization method has been proposed for the eigenstates of nonrelativistic QBs, whose shapes are given by a $w(z)$ with the properties~\refeq{WZ}. These are governed by the nonrelativistic Schr\"odinger equation~\refeq{Schr} subject to the Dirichlet BC. The ansatz for the eigenfunctions is given in terms of a linear combination of the orthogonal eigenfunctions of the circular QB. For a circular billiard with radius $r_0$ the domain is described by~\refeq{WZ} with $w(z)=z$, $\vert w(z)\vert =r$, $e^{i\theta(z)}=e^{i\varphi}$, and at the boundary $w(\varphi)=r_0e^{i\varphi}$ and $e^{i\alpha(\varphi)}=e^{i\varphi}$. The eigenvalues $\epsilon_n$ are given as the zeros of the Bessel functions and the eigenfunctions $\psi_n(r,\varphi)$ are given in terms of the Bessel functions,
\be
J_n(\epsilon_{n,\nu}r_0)=0,\, \psi^{e,o}_{n,\nu}(r,\varphi)=J_{n,\nu}(\epsilon_{n,\nu}r)f^{e,o}_n(\varphi), 
\ee 
with $\nu=1,2,\dots,n=1,2,\dots$, $f^e_n(\varphi)=\cos(n\varphi)$ and $f^o_n(\varphi)=\sin(n\varphi)$ for solutions that are symmetric and antisymmetric with respect to the $x$ axis, respectively. The index $\nu$ denotes the $\nu$th zero of $J_n(\epsilon_{n,\nu}r_0)$ for a given index $n$. The ansatz used in~\cite{Robnik1984,Berry1986} for the solutions of the Schr\"odinger equation~\refeq{Schr} with Dirichlet BCs along the boundary $w(\varphi)$ reads
\be
\label{ansatzpsi}
\psi(r,\varphi)=\sum_{n=-\infty}^\infty\sum_{\nu=1}^\infty \frac{c_{n,\nu}(k)N_{n,\nu}}{\epsilon_{n,\nu}}J_{\vert n\vert,\nu}(\epsilon_{n,\nu}r)e^{in\varphi},
\ee
with $N^{-1}_{n\nu}=\sqrt{\pi}\vert J^\prime_{n,\nu}\vert$, $c_{n,\nu}$ denoting the expansion coefficients and $\epsilon_{n,\nu}=\epsilon_{-n,\nu}$. These wave functions fulfill the Dirichlet BCs by construction. The eigenvalues and associated eigenstates are obtained by solving the eigenvalue problem
\be
\mathcal{\hat M}\boldsymbol{c}=\frac{1}{k^2}\boldsymbol{c},
\label{CMMQB1}
\ee
with matrix elements $\mathcal{N}_{n\nu m\mu}\mathcal{\hat M}_{n\nu m\mu}$ given by
\be
\int_0^{r_0}rdr\int_0^{2\pi}d\varphi\vert w^\prime (z)\vert^2 e^{i(n-m)\varphi}J_{\vert n\vert,\nu}(\epsilon_{n,\nu}r)J_{\vert m\vert,\nu}(\epsilon_{m,\mu}r),
\label{CMMQB2}
\ee
where $\mathcal{N}_{n\nu m\mu}^{-1}=\frac{N_{n\nu} N_{m\mu}}{\epsilon_{n,\nu}\epsilon_{m,\mu}}$.
In~\refsec{Numerics} we demonstrate the efficiency of this method for the Africa QB~\cite{Berry1986}. 

\section{The flaws of the CMM for neutrino billiards\label{Proof}}
In analogy to the quantization procedure~\cite{Robnik1984,Berry1986}, in Ref.~\cite{Xu2013} a method was proposed for the computation of the eigenstates of NBs on the basis on those of a circular NB. The eigenvalues and eigenfunctions of the circular NB are obtained by imposing the BC~\refeq{BC2} on the ansatz~\refeq{Phi1} with the second component given in~\refeq{Phi2}, leading to the quantization condition~\cite{Berry1987}
\be
\label{BC5}
J_{m+1}(\kappa_{m,\mu})=J_m(\kappa_{m,\mu}),
\ee 
where $\mu$ counts the number of eigenstates below $k_{m,\mu}$ for a given total angular momentum $m+\frac{1}{2}$, 
that is, the $\kappa_{m,\mu}$ are sorted as $\kappa_{m,1}\leq\kappa_{m,2}\leq\dots$. The associated eigenfunctions read
\ba
{\Phi_1}_{m,\mu}(r,\varphi)\label{eigenfctcirc}
&=&\mathcal{N}_{m,\mu}i^mJ_m(\kappa_{m,\mu}r)e^{im\varphi},\\
{\Phi_2}_{m,\mu}(r,\varphi)\nonumber
&=&\mathcal{N}_{m,\mu}i^{m+1}J_{m+1}(\kappa_{m,\mu}r)e^{i(m+1)\varphi}.
\ea
The spinor eigenfunctions 
\be
\boldsymbol{\Phi}_{m,\mu}(r,\varphi)=\begin{pmatrix}
	{\Phi_1}_{m,\mu}(r,\varphi) \\ {\Phi_2}_{m,\mu}(r,\varphi)
\end{pmatrix}
\ee
are orthogonal to each other~\cite{Berry1987} [cf.~\refsec{suppl}], 
\be
\label{orth}
\int_0^{r_0}rdr\int_0^{2\pi}d\varphi\boldsymbol{\Phi}_{m,\mu}(r,\varphi)\cdot \boldsymbol{\Phi}_{n,\nu}(r,\varphi) =\delta_{n,m}\delta_{\nu,\mu}.
\ee
In analogy to Ref.~\cite{Robnik1984} for the CMM the ansatz for the first spinor component consists of a superposition of the eigenfunctions of the circular NB,
\ba
\label{ansatzpsi1}
&&\psi_1(r,\varphi)=\\
&&\sum_{n=-\infty}^\infty\sum_{\nu=1}^\infty a_{n,\nu}(k)i^nJ_n(r\kappa_{n,\nu})e^{in\varphi}
=\sum_ja_j(k){\Phi_1}_j(r,\varphi),\nonumber
\ea
where the index $j$ counts the number of eigenvalues below $\kappa_{l(j),\lambda(j)}$, that is, these are sorted by size, $\kappa_{l(1),\lambda(1)}\leq \kappa_{l(2),\lambda(2)}\leq \dots$. Note, that for circular NBs we have $\kappa_{n,\nu}\ne\kappa_{-n,\nu}$~\cite{Berry1987}. 

The ansatz for the second component, $\psi_2(r,\varphi)$ is obtained by inserting $\psi_1(r,\varphi)$ into the Dirac equation~(\ref{DGLpsi2}), from which the associated Schr\"odinger equations for the two spinor components, that are solved within the CMM, originate, yielding
\ba
\label{ansatzpsi2}
&&[w^\prime(z)]^\ast\psi_2(r,\varphi)=\\
&&\sum_{n,\nu} b_{n,\nu}(k)i^{n+1}J_{n+1}(r\kappa_{n,\nu})e^{i(n+1)\varphi}
=\sum_jb_j(k){\Phi_2}_j(r,\varphi),\nonumber\\ 
&&b_{n,\nu}(k)=\frac{\kappa_{n,\nu}}{k}a_{n,\nu}(k).\nonumber
\label{bln}
\ea
Hence, for $k\ne\kappa_{n,\nu}$ and $[w^\prime(z)]^\ast\ne 1$, that is, for shapes different from a circle we have $\frac{1}{[w^\prime(z)]^\ast}b_{n,\nu}\ne a_{n,\nu}$, in contrast to the assumptions of~\cite{Xu2013}. Inserting this result into~\refeq{DGLpsi1} yields for $\psi_1(r,\varphi)$ the Schr\"odinger equation
\begin{align}
\label{Schrpsi1}
&\sum_{n,\nu}a_{n,\nu}(k)\kappa^2_{n,\nu}i^nJ_n(\kappa_{n,\nu}r)e^{in\varphi}=\\
&k^2\vert w^\prime(z)\vert^2\sum_{n,\nu} a_{n,\nu}(k)i^nJ_n(\kappa_{n,\nu}r)e^{in\varphi}.
\nonumber\end{align}
Similarly, inserting~\refeq{ansatzpsi2} into~\refeq{Schr} for $i=2$ gives 
\ba
\label{Schrpsi2a}
&&r^2\Delta_{(r,\varphi)}\psi_2(r,\varphi)=\\
&&\left(r\frac{\partial}{\partial r} +i\frac{\partial}{\partial \varphi}\right)\left[\frac{iz}{k[w^\prime(z)]^\ast}\sum_{n,\nu} a_{n,\nu}(k)\kappa^2_{n,\nu}i^{n}J_{n}(r\kappa_{n,\nu})e^{in\varphi}\right].
\nonumber\ea
Assuming that the $k$ value is chosen such that the condition~\refeq{Schrpsi1} holds to replace the sum over $n,\nu$ by the right-hand side of this equation, yields  
\ba
\label{Schrpsi2}
&&\sum_{n,\nu} b_{n,\nu}(k)\kappa^2_{n,\nu}i^{n+1}J_{n+1}(\kappa_{n,\nu}r)e^{i(n+1)\varphi}=\\
&&k^2\vert w^\prime(z)\vert^2\sum_{n,\nu} b_{n,\nu}(k)i^{n+1}J_{n+1}(\kappa_{n,\nu}r)e^{i(n+1)\varphi}+\delta,
\nonumber\ea
with $\delta=0$. Yet, when this equation and~\refeq{Schrpsi1} do not have solutions at the same values of $k$, then an additional term
\be
\label{delta}
\delta =2i kw^\prime(z)[w^{\prime\prime}(z)]^\ast\psi_1(r,\varphi)
\ee
appears in~\refeq{Schrpsi2}.
Note, that these equations are no identities, but conditional equations for the eigenstates of the associated NB. 

Defining in analogy to the nonrelativistic case,~\refeq{CMMQB2}, matrices $\mathcal{\hat K}_1$, $\mathcal{\hat K}_2$, $\mathcal{\hat J}_1$ and $\mathcal{\hat J}_2$, [cf.~Eq.~(S11) in~\refsec{suppl}], Eqs.~(\ref{Schrpsi1}) and~(\ref{Schrpsi2}) can be written as
\ba
&&\left[\mathcal{\hat K}_1-k^2\mathcal{\hat J}_1\right]\boldsymbol{a}=0,\label{Schr2psi1}\\
&&\left[\mathcal{\hat K}_2-k^2\mathcal{\hat J}_2\right]\boldsymbol{b}=0.\label{Schr2psi2}
\ea
They have solutions only for discrete values of $k=\tilde k_j$, because the matrices $\mathcal{\hat K}_1,\mathcal{\hat K}_2,\mathcal{\hat J}_1,\mathcal{\hat J}_2$ do not depend on $k$. 

In~\cite{Xu2013} $\boldsymbol{b}$ entering~\refeq{ansatzpsi2} is set equal to $\boldsymbol{a}$ in~\refeq{ansatzpsi1}, and the eigenstates $[\boldsymbol{a}_j(\tilde k_j), \tilde k_j]$ of the sum of Eqs.~(\ref{Schr2psi1}) and~(\ref{Schr2psi2}), $\left[\left(\mathcal{\hat K}_1+\mathcal{\hat K}_2\right)-\tilde k_j^2\left(\mathcal{\hat J}_1+\mathcal{\hat J}_2\right)\right]\boldsymbol{a}_j=0$, are determined using the orthogonality~\refeq{orth}, 
\ba
\label{Solaa}
&&\mathcal{\hat J}-\frac{\hat\kappa^2}{k^2}=\hat 0,\, \mathcal{\hat J}_{jj^\prime}=\left(\mathcal{M}_a^T\left[\mathcal{\hat J}_1+\mathcal{\hat J}_2\right]\mathcal{M}_a\right)_{jj^\prime},\\
&&\hat\kappa^2_{jj^\prime} =\left(\mathcal{M}_a^T\left[\mathcal{\hat K}_1+\mathcal{\hat K}_2\right]\mathcal{M}_a\right)_{jj^\prime}=\kappa^2_j\delta_{jj^\prime},
\nonumber\ea
where the column vectors of the matrix $\mathcal{M}_a$ are the eigenvectors $\boldsymbol{a}_j$ corresponding to the eigenvalues $\tilde k_j,j=1,\dots , N$ and the dimension $N$ equals the number of eigenstates of the circular NB, $\kappa_1\leq\kappa_2\leq\dots\leq\kappa_N$, taken into account. Yet, for $\boldsymbol{b}=\boldsymbol{a}$ the ansatz for the spinor functions with components~\refeq{ansatzpsi1} and~\refeq{ansatzpsi2} is no solution of the Dirac equation~(\ref{DGLpsi1}) with~(\ref{DGLpsi2}). In fact, Eqs.~(\ref{Schr2psi1}),~(\ref{Schr2psi2}) for the spinor components and~(\ref{Solaa}) do not have common solutions and the additional term $\delta$ in~\refeq{Schrpsi2a} is nonzero, except for constant $w^\prime (z)$, that is, for circular shapes. The reason for this deficiency is that, as mentioned above, Eqs.~(\ref{Schrpsi1}) and~(\ref{Schrpsi2}) are no identities, as in the case of the  ansatz~\refeq{Phi1} with~\refeq{Phi2}, but conditional equations. 

The correct strategy to solve the Dirac equation~(\ref{DGLpsi1}) with~(\ref{DGLpsi2}) using the ansatz~\refeq{ansatzpsi1}, would be to solve the generalized eigenvalue problem~\refeq{Schr2psi1} for the first component, because for $\boldsymbol{a}\ne\boldsymbol{b}$ the orthogonality of the eigenstates of the circular NB cannot be employed. Then the second component would be obtained from~\refeq{ansatzpsi2}. Again,~\refeq{Schr2psi1} has solutions only for discrete values of $k$, since the matrices $\mathcal{\hat J}_i,\mathcal{\hat K}_i,i=1,2$ are assumed to be independent of $k$. However, like for the case $\boldsymbol{a} =\boldsymbol{b}$ these differ from those of~\refeq{Schr2psi2} for non-circular shapes and $\delta\ne 0$ in~\refeq{Schrpsi2a}; cf.~\refsec{Numerics} and~\refsec{Example2}

By construction, the solutions of~\refeq{Solaa} ($\tilde k_j,{\boldsymbol\psi}_j(r,\varphi)$) fulfill the BC for circular NBs, ${\psi_1}_j(r_0,\varphi)=-ie^{-i\varphi}{\psi_2}_j(r_0,\varphi)$, that is,
\be
\label{BC6a}
{\psi_1}_j(r_0,\varphi)=i^{-1}e^{-i\varphi}\sum_{n,\nu} a_{n,\nu}(\tilde k_j)i^{n+1}J_{n+1}(r_0\kappa_{n,\nu})e^{i(n+1)\varphi}.
\ee
Yet, they fail to comply with the BC for NBs,~\refeq{BC2}, 
\be
{\psi_1}_j(r_0,\varphi)\ne\frac{1}{\vert w^\prime(\varphi)\vert}\sum_{n,\nu} b_{n,\nu}(\tilde k_j)i^{n+1}J_{n+1}(r_0\kappa_{n,\nu})e^{i(n+1)\varphi},
\label{BC6b}
\ee
also when using the correct ansatz~\refeq{ansatzpsi1} with~\refeq{ansatzpsi2} for the solutions of the Dirac equation(\ref{DGLpsi1}) with~(\ref{DGLpsi2}). Indeed, the right-hand sides of equations~(\ref{BC6a}) and~(\ref{BC6b}) are only equal, that is, the BC~\refeq{BC2} only holds for $w^\prime(\varphi)=ie^{i\varphi}, \tilde k_j=\kappa_{n,\nu}$, implying that the outgoing flux
\ba
\label{CurrCMM}
\boldsymbol{n}\cdot\left[\boldsymbol{\psi}^\dagger\boldsymbol{\hat\sigma}\boldsymbol{\psi}\right]&&=\cos[\alpha(\varphi)]\Re[\psi^\ast_1\psi_2]+\sin[\alpha(\varphi)]\Im[\psi^\ast_1\psi_2]\nonumber\\
&&\equiv\sin[\alpha(\varphi)-\varphi]\vert\psi_1(\varphi)\vert^2,
\ea
is nonvanishing for non-circular shapes and proportional to the size of the intensity of the wave function $\psi_1(\varphi)$ along the boundary. 

The impact of the boundary wave functions $\psi_1(\varphi)$ and $\psi_2(\varphi)$ remains non-negligible in the semiclassical limit $k\to\infty$~\cite{Dietz2020}. This can be deduced from the feature of the trace formula, which is the semiclassical approximation for the spectral density in terms of a sum over periodic orbits of the classical billiard of corresponding shape~\cite{Gutzwiller1971}. The difference between the trace formulas for QBs and NBs is, that in the latter contributions of periodic orbits with an odd number of reflections at the boundary are missing, because the contributions from the spinor components of its eigenfunctions cancel each other~\cite{Berry1987,Bolte1999,Wurm2009,Dietz2020}, which would not be possible if they vanished along the boundary like for QBs. Note, that this feature of the trace formula is not a characteristic of relativistic QBs, but it is also observed in non-relativistic systems subject to a vectorial wave equation~\cite{Balian1977,Dembowski2002}.

Still, the fluctuation properties of the eigenvalues obtained from each of the equations~(\ref{Schr2psi1}),~(\ref{Schr2psi2}) and~(\ref{Solaa}) are expected to be close to those of the NB. Namely, like the eigenfunctions of the circular NB, those of these equations are complex, that is they are not invariant under time-reversal. Furthermore, the potential $\vert w^\prime(z)\vert^2$ induces the same chaoticity as for the CMM for nonrelativistic QBs. Accordingly, the spectral properties may not be used as a criterion to prove or disprove the applicability of the CMM for NBs. To illustrate in the Secs.~\ref{Numerics} and~\ref{Example2} numerically for a few examples the failure of the CMM for NBs, we consider the Weyl formula for the smooth part of the integrated spectral density, which differs for NBs and QBs~\cite{Weyl1929,Berry1987}, and demonstrate that that for QBs applies for CMM, whereas for the eigenvalues obtained from BIM excellent agreement with that for NBs is found. Furthermore we compute differences between the eigenvalues obtained with equations~(\ref{Schr2psi1}),~(\ref{Schr2psi2}), and from the BIM and~\refeq{Solaa}, respectively, and compare wave functions obtained with BIM and CMM.

We conclude that the method proposed in~\cite{Xu2013,Huang2018} for the determination of the eigenstates of NBs is only applicable to the circular NB, that is for the trivial case. Otherwise, it does not provide reliable solutions, because these do not comply with the Dirac equation and BCs for neutrino billiards. It is stated in these works that the solutions of the CMM fulfill the BC~\refeq{BC2}. Yet, for this the second spinor component $\psi_2(r,\varphi)$,~\refeq{ansatzpsi2} needs to be multiplied with a $\varphi$-dependent factor, which can be read off Eqs.~(\ref{BC6a}) and~(\ref{BC6b}). However then, even when using the correct ansatz for the spinor components,~\refeq{ansatzpsi1} and~\refeq{ansatzpsi2}, the Dirac equation is not fulfilled since application of the operator~\refeq{Grad} to this factor is nonvanishing. This reflects the fact, that it is not sufficient to solve the Schr\"odinger equation for the spinor components, as demonstrated, e.g., for the ellipse NB in~\cite{Dietz2019}. 

In Ref.~\cite{Song2019} the authors apply the same method to massive NBs with mass $m_0\ne 0$. We demonstrate in Ref.~\cite{Dietz2020}, that the corresponding Dirac equation can be brought to the same form as in~(\ref{DGLpsi1}) and~(\ref{DGLpsi2}) with modified BCs that comply with the ultrarelativistic limit $m_0\to 0$ for massless NBs and the nonrelativistic limit, $\frac{m_0c}{\hbar k}\to\infty$. Starting from this Dirac equation with modified BCs and proceeding as above, it can be shown, that the method of~~\cite{Song2019} does not provide the eigenstates of NBs with non-circular shapes. 

In the Secs.~\ref{Numerics} and~\ref{Example2} we provide examples, which show that the eigenvalues obtained from~\refeq{Solaa} can be close to those of the corresponding NB, for shapes with nearly constant curvature except in small parts of the boundary, however not the eigenfunctions. So as long as one is not interested in the eigenfunctions and their phases, and the shape is close to that of a circle, it can serve as an approximate method to determine a few 1000th of eigenvalues. On the other hand, the BIM introduced in Ref.~\cite{Berry1987} for the ultrarelativistic case and its extension to massive NBs~\cite{Dietz2020,Dietz2022a} yields at least that many eigenvalues with low numerical effort -- especially for the NBs under consideration -- and also the associated eigenfunctions with high precision. Thus, there is no reason to use a method, which has been proven to be erroneous and thus provides unreliable eigenstates. The flaws of the CMM unveiled in this work, in fact explain the discrepancies between the solutions of the CMM and expected results observed, e.g. in Refs.~\cite{Song2019,Li2022}. 

\section{Numerical analysis of the CMM for billiards from the family of the Africa billiard for which the eigenvalues are close to those of the NB\label{Numerics}}
\begin{figure}[!h]
\centering
\includegraphics[width=0.495\linewidth]{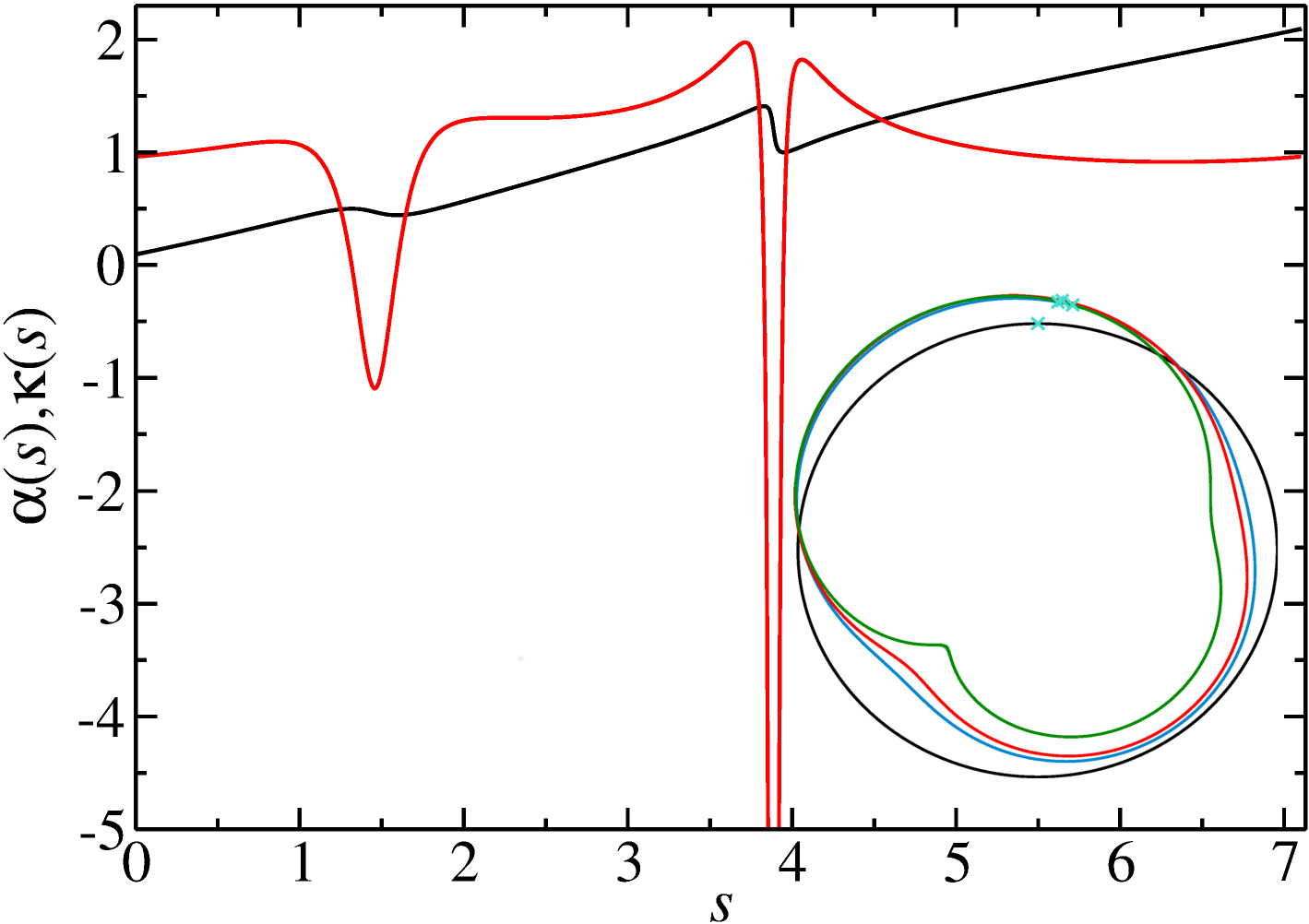}
\includegraphics[width=0.485\linewidth]{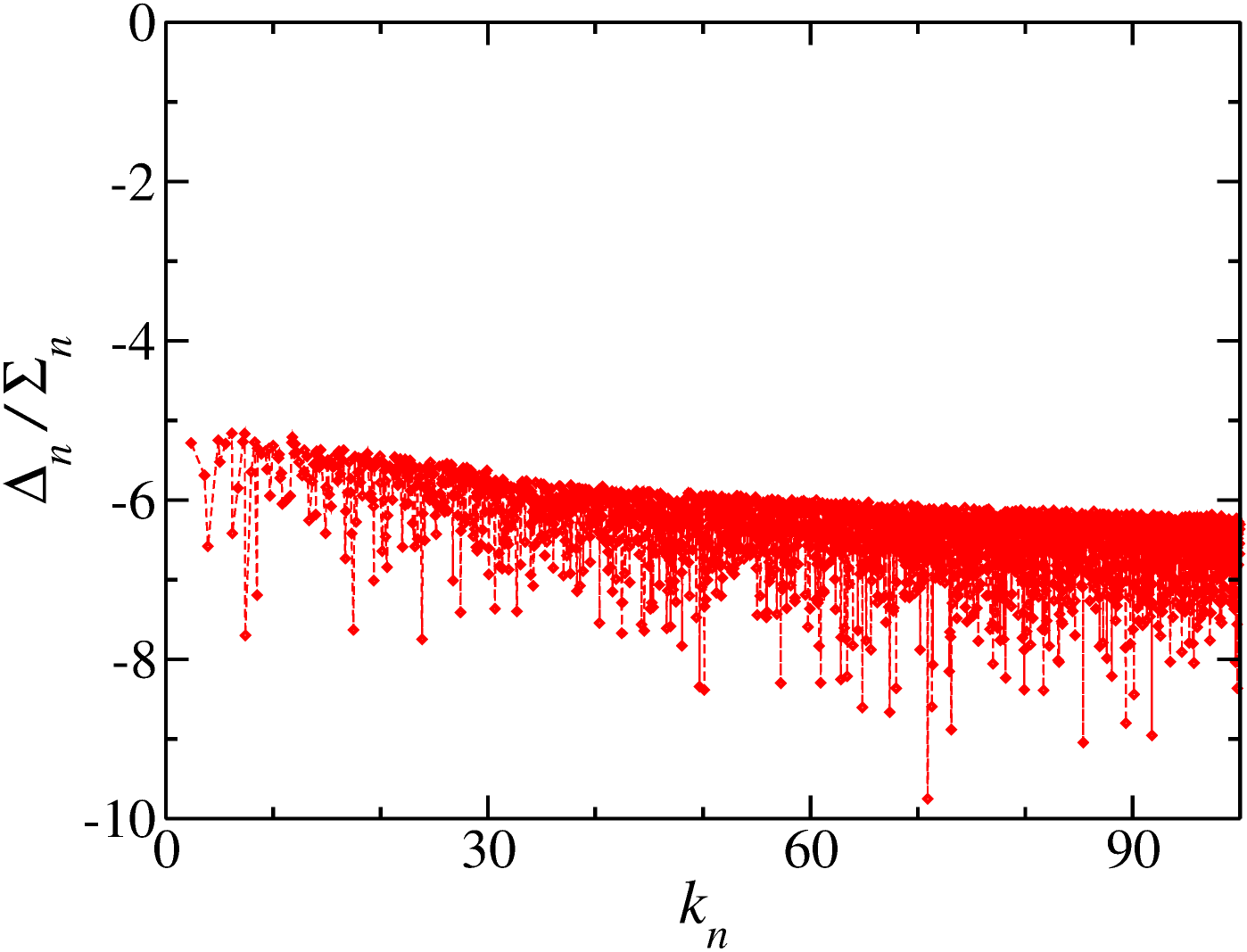}
	\caption{Left: The angle $\alpha(s)$ of the normal vector with respect to the $x$ axis (black) and boundary curvature (red) for the shape of the Africa billiard defined in~\refeq{Shape} with $\alpha=0.2$, where $s$ denotes the arclength parameter. The shapes considered in this work are shown in the inset for $\alpha=0$ (black), $\alpha=0.1$ (blue), $\alpha=0.125$ (red) and $\alpha=0.2$ (green). The crosses mark the zero point of the arclength parameter $s$, which increases along the boundary in clockwise direction. Right: Relative differences $\Delta_n/\Sigma_n=2\frac{|k^{BIM}_n-k^{CMM}_n|}{k^{BIM}_n+k^{CMM}_n}$ for the Africa QB with $\alpha=0.2$. Here, $k_n^{BIM}$ and $k_n^{CMM}$ denote the eigenvalues obtained with the BIM and CMM, respectively.}
\label{Fig1}
\end{figure}

We compare results obtained with the CMM and the BIM, respectively, for a family of billiards with shapes given as~\cite{Berry1986,Berry1987}
\be
\label{Shape}
w^{AF}(z;\alpha)=\frac{z+\alpha z^2+\alpha e^{i\pi/3}z^3}{\sqrt{1+5\alpha^2}}.
\ee
For $\alpha=0$ the shape is circular, whereas with increasing $\alpha$ the classical dynamics undergoes a transition from regular via mixed regular-chaotic to chaotic. We computed eigenstates for several values of $\alpha\leq 0.2$. In the inset of the left part of~\reffig{Fig1} shapes are shown for $\alpha=0$ (black), and for the cases considered, $\alpha=0.1$ (blue), $\alpha=0.125$ (red) and $\alpha=0.2$. For the latter one, known as 'Africa billiard', the dynamics is chaotic. Furthermore, the angle $\alpha(s)$ of the normal vector with respect to the x-axis and the curvature $\kappa(s)$ are exhibited as function of the arclength parameter. The angle $\alpha(s)$ varies nearly linearly with $s$ and the curvature is close to unity, i.e. to that of a circle of radius  unity, except in the regions around the two bulges.

To verify correctness of the codes used to solve Eqs.~(\ref{Schr2psi1}),~(\ref{Schr2psi2}) and~(\ref{Solaa}), we first applied the CMM~\refeq{CMMQB1} with~\refeq{CMMQB2} to the corresponding QBs, as the associated eigenvalue equations are similar to these equations. For the BIM the eigenvalues $k_n$ and eigenfunctions are determined by solving a quantization condition of the form $\det\hat A(k_n)=0$, where the matrix $\hat A$ is obtained from a boundary-integral equation deduced from the Green theorem~\cite{Berry1987,Baecker2003,Dietz2020}. In the examples presented in the following we chose for the QB and NB the same dimension $500\times 500$ for $\hat A$, which corresponds to the discretization of the boundary into 500 pieces, and computed it for 600000 values of $k$ in the range of $k\in [0,100]$ to obtain complete sequences of $\approx 3000$ eigenvalues. 
For the CMM~\refeq{ansatzpsi} we considered 5000 eigenstates of the circular QB, yielding $\simeq 2500$ reliable eigenvalues, as demonstrated in~\reffig{Fig1}. There we show the relative deviations of the eigenvalues resulting from the BIM and from the CMM, respectively. 

\begin{figure}[!h]
\centering
\includegraphics[width=0.5\linewidth]{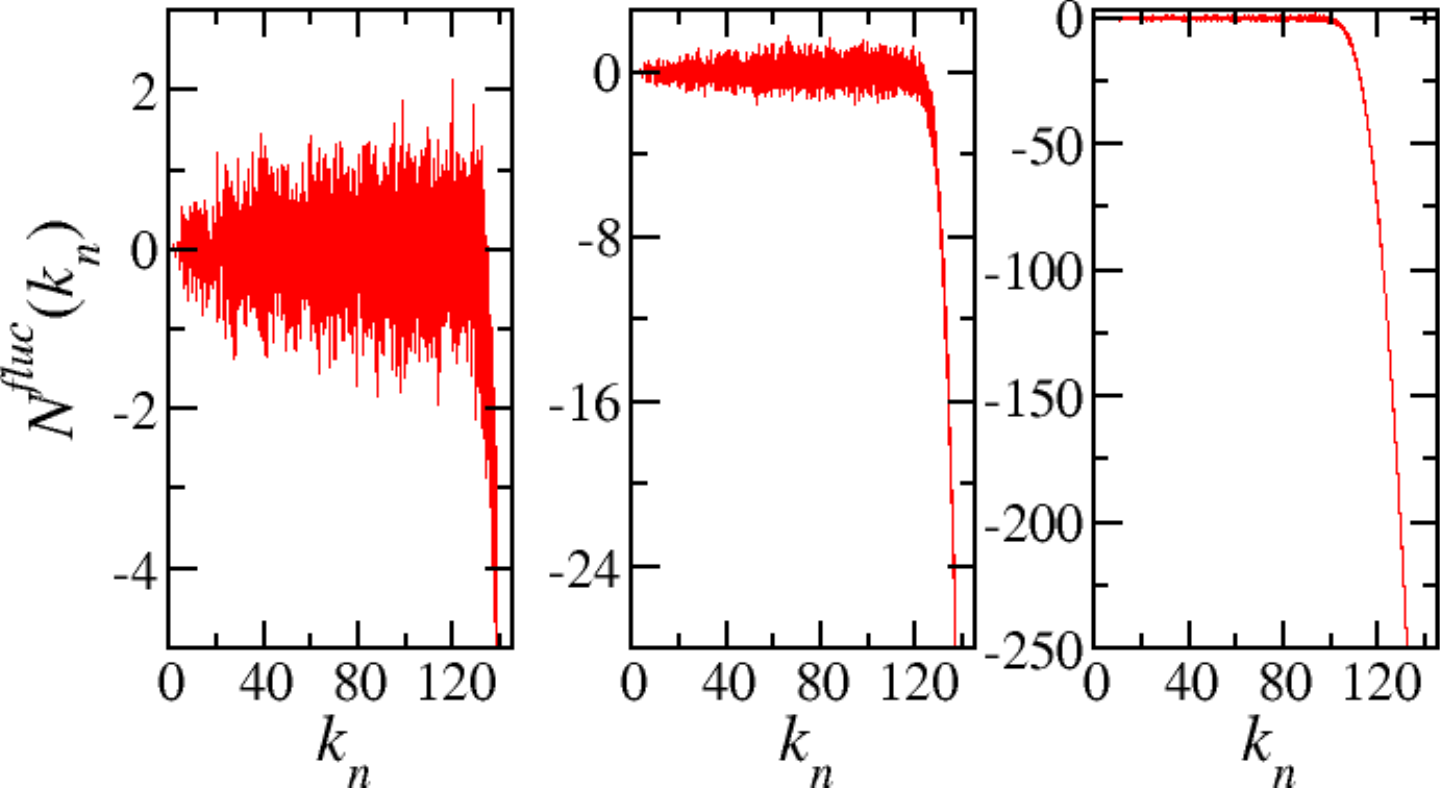}
\includegraphics[width=0.44\linewidth]{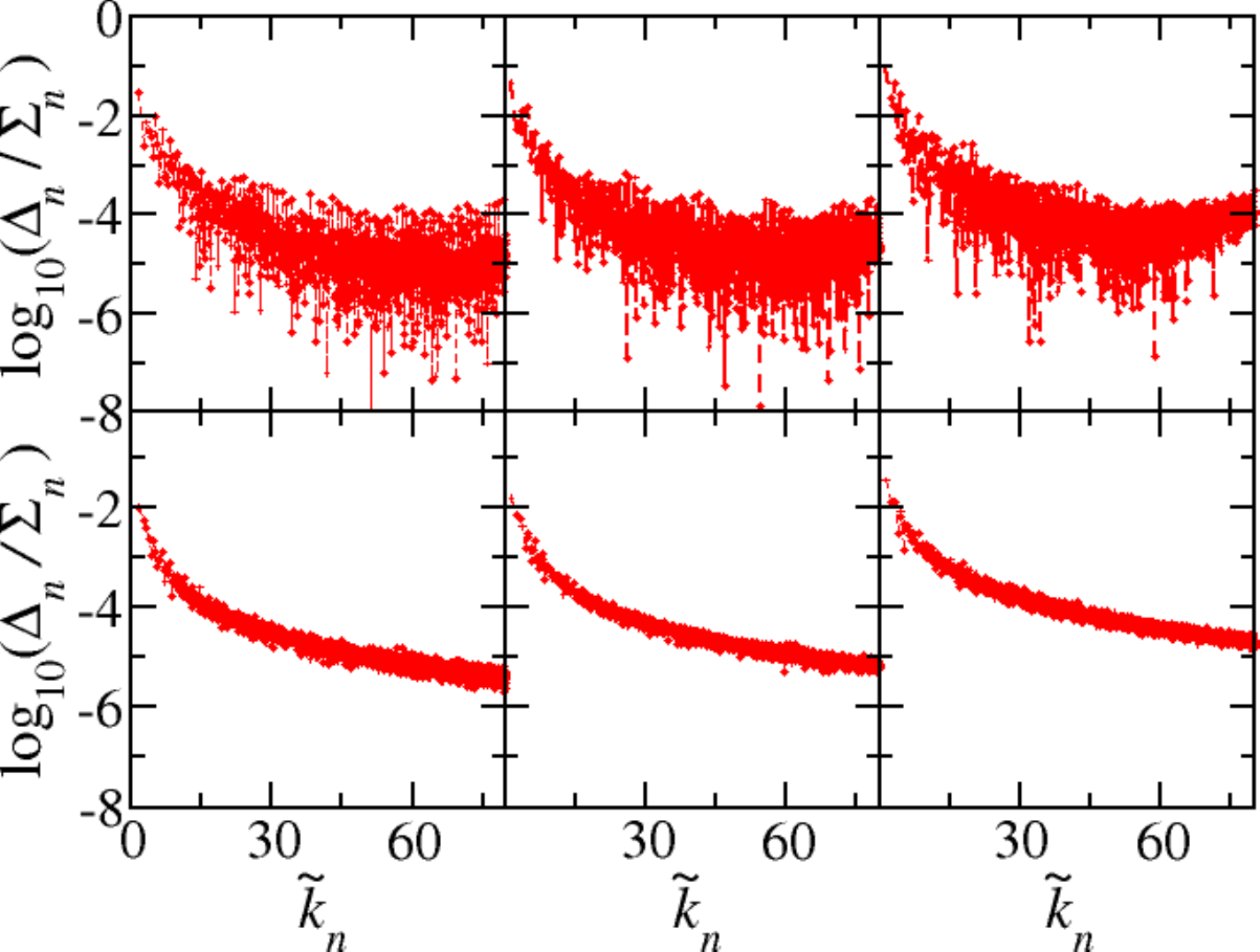}
\caption{
	Left: Difference $N^{fluc}(k_n)=N(k_n)-N^{Weyl}(k_n)$ between the number of eigenvalues $N(k_n)$ of~\refeq{Solaa} below $k=k_n$ and the Weyl formula $N^{Weyl}(k_n)=\frac{\mathcal{A}}{4\pi}k_n^2+C$ for an NB with shape~\refeq{Shape} with $\alpha=0.1$ (left), $\alpha=0.125$ (middle) and $\alpha=0.2$ (right). Right: Relative differences $\Delta_n/\Sigma_n=2\frac{|k^a_n-k^b_n|}{k^a_n+k^b_n}$ for $\alpha=0.1$ (left), $\alpha=0.125$ (middle) and $\alpha=0.2$ (right). In the upper panels the $k^a_n$ are obtained from~\refeq{Schr2psi1} and the $k^b_n$ are deduced from~(\ref{Schr2psi2}), in the lower ones the $k^a_n$ were computed with the BIM and the $k^b_n$ are obtained from~\refeq{Solaa}.} 
\label{Fig2}
\end{figure}

For the NB we used the first 10000 eigenstates of the circular NB to compute the solutions of \refeq{Schr2psi1},~\refeq{Schr2psi2} and~\refeq{Solaa}. The number of eigenstates obtained from the CMM is of the same order as for the QB, even though we considered there only half the number of eigenstates of the circular QB. Beyond that value, the Weyl formula for NBs~\cite{Berry1987}, which gives the average of the number of eigenvalues $N(k_n)$ below $k=k_n$, $N^{Weyl}(k_n)=\frac{\mathcal{A}}{4\pi}k_n^2+C$ with $\mathcal{A}$ and $C$ denoting the area of the corresponding NB and a positive constant $C\lesssim 0.5$, is no longer applicable [cf. the left part of~\reffig{Fig2}]. In fact, to verify whether the eigenvalues obtained from the CMM are those of NBs, we fit a polynomial of the form $N^{smooth}(k_n)=\frac{\mathcal{\tilde A}}{4\pi}k_n^2-\frac{\mathcal{\tilde L}}{4\pi}k_n+C$ to $N(k_n)$, which actually is the Weyl formula for non-relativistic QBs with area $\mathcal{\tilde A}$ and perimeter $\mathcal{\tilde L}$. For the BIM the boundary term $\frac{\mathcal{\tilde L}}{4\pi}$ is by a factor of $10^{-5}-10^{-4}$ smaller than the area term, whereas for the solutions of Eqs.~(\ref{Schr2psi2}) and~(\ref{Solaa}) the factor increases from $\approx 10^{-3}$ to $\approx 10^{-2}$ with increasing deformation of the shape from that of a circle, and for the solutions of~\refeq{Schr2psi1} this factor is about 10 times smaller than these values. Accordingly, for the CMM the parameter $\mathcal{\tilde A}$ deviates from the expected value, $\mathcal{A}$. 

In the lower panels of the right part of~\reffig{Fig2} we show the relative deviations of the eigenvalues obtained with the BIM from those computed with~\refeq{Solaa}, in the upper panels those between the eigenvalues obtained from Eqs.~(\ref{Schr2psi1}) and~(\ref{Schr2psi2}), respectively. The distances are of similar size for all cases and by at least two decades larger than for the nonrelativistic case, thus confirming that the CMM does not yield solutions of the Dirac equation for NBs. Still, the agreement with the eigenvalues obtained from the exact quantization procedure, i.e., the BIM is good, especially for the solutions of~\refeq{Schr2psi1}, even for the case with $\alpha=0.2$. This maybe explained by the fact that the boundary curvature is close to that of the circle except in the region around the bulges. However, generally, as demonstrated in~\reffig{Fig2}, the deviations between the eigenvalues obtained with the BIM and CMM increase with the deformation of the circular NB, thus corroborating the proof. 

Furthermore, the boundary wave functions for the BIM and CMM cases, show clear differences. In Fig.~S1 of~\refsec{suppl} the outgoing flux is shown for two examples. It is of the order of $10^{-16}$ for the BIM results and agrees well with the analytical result~\refeq{CurrCMM} for the CMM ones. A clear discrepancy is visible for all cases also in the corresponding wave functions shown in Fig.~S2 of~\refsec{suppl}, one reason being that their phases differ due to the differing boundary conditions. The wave functions for state 2508 are scarred along a straight line corresponding to a remnant of the diameter orbit in the original circular NB. It bounces back and forth between two nearly circular focusing boundary segments, and thus is expected to occur in both cases, with dominant support in the region where the boundary curvature is close to unity. Indeed, such scarred wave functions are also observed in the Africa QB. Nevertheless, to get for the CMM case the correct phase along that orbit, the wave function has to be multiplied with a factor in order to fulfill the BC for the NB [cf.~\refeq{BC6b}]. 

\subsection{Example of a billiard for which for the CMM yields solutions that clearly deviate from those of the BIM\label{Example2}}
\begin{figure}[!h]
\centering
\includegraphics[width=0.483\linewidth]{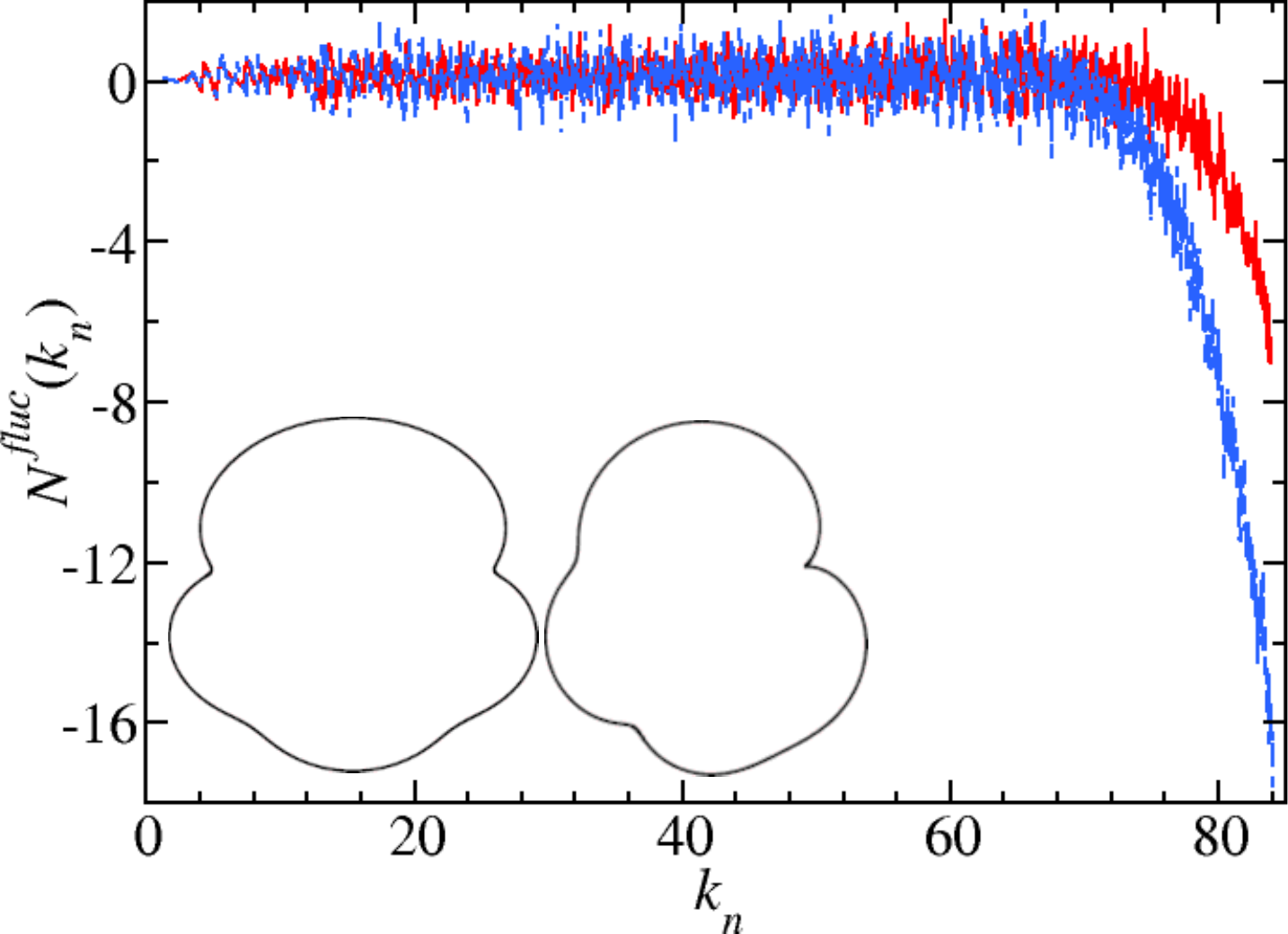}
\includegraphics[width=0.497\linewidth]{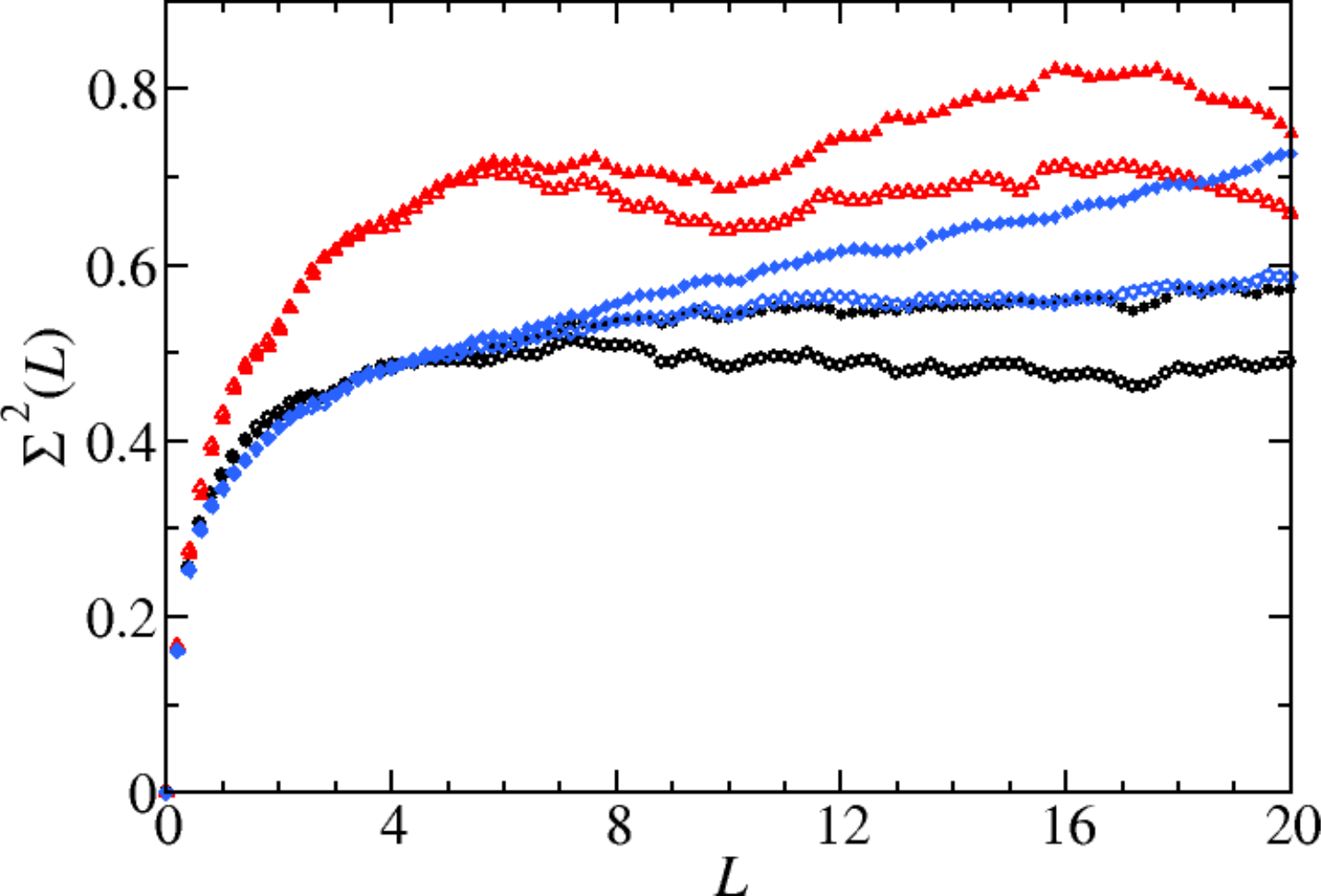}
\caption{
        Left: Difference $N^{fluc}(k_n)=N(k_n)-N^{smooth}(k_n)$ between the number of eigenvalues obtained from~\refeq{Solaa}, $N(k_n)$, below $k=k_n$ and the best fitting Weyl formula for QBs, $N^{smooth}(k_n)=\frac{\mathcal{\tilde A}}{4\pi}k_n^2-\frac{\mathcal{\tilde L}}{4\pi}k_n+C$ for the shape with real (red) and complex (blue) $c_3$ [cf. insets].
	Right: Comparison of the number variance $\Sigma^2(L)$ computed for the eigenvalues of the NB using BIM (open symbols) and obtained from~\refeq{Solaa} for the billiards with real (red) and complex (black) $c_3$ and for the Africa shape with $\alpha =0.2$ (blue), respectively. Here, the eigenvalues were unfolded with the Weyl formula for nonrelativistic QBs (full symbols), taking into account as many eigenvalues as possible, that is, ensuring that none are missing according to the unfolding procedure.
}
\label{Fig02}
\end{figure}

We also considered two billiards with shapes given by~\refeq{WZ} with $c_1=1$, $c_2=c_4=c_5=0.1$, and $c_3=0.1$ -- exhibiting a mirror symmetry --, respectively, $c_3=0.1e^{i\pi/3}$ -- deviating along most parts of the boundary from that of a circle. The shapes are shown in the left and right inset of the left part of~\reffig{Fig02}, respectively. To obtain the solutions of  Eqs.~(\ref{Schr2psi1}),~(\ref{Schr2psi2}) and~(\ref{Solaa}) we used the same number of eigenvalues of the corresponding circular QB and NB as in the previous examples. However, in that case only about 1000 eigenvalues could be obtained that comply with the Weyl formula in the sense that the smooth part of the integrated spectral density is well described by that for nonrelativistic QBs, that is, a second-order polynomial [cf. left part of~\reffig{Fig02}]. Namely $\mathcal{\tilde L}$ is by a factor $10^{-5}$ smaller than the area term for the BIM, whereas for the solutions of Eqs.~(\ref{Schr2psi2}) and~(\ref{Solaa}) this factor is $\approx 10^{-2}$ and $\approx 10^{-1}$ for complex and real $c_3$, respectively, and for the solutions of~\refeq{Schr2psi1} about 10 times smaller than these values. 

\begin{figure}[!h]
\centering
\includegraphics[width=0.49\linewidth]{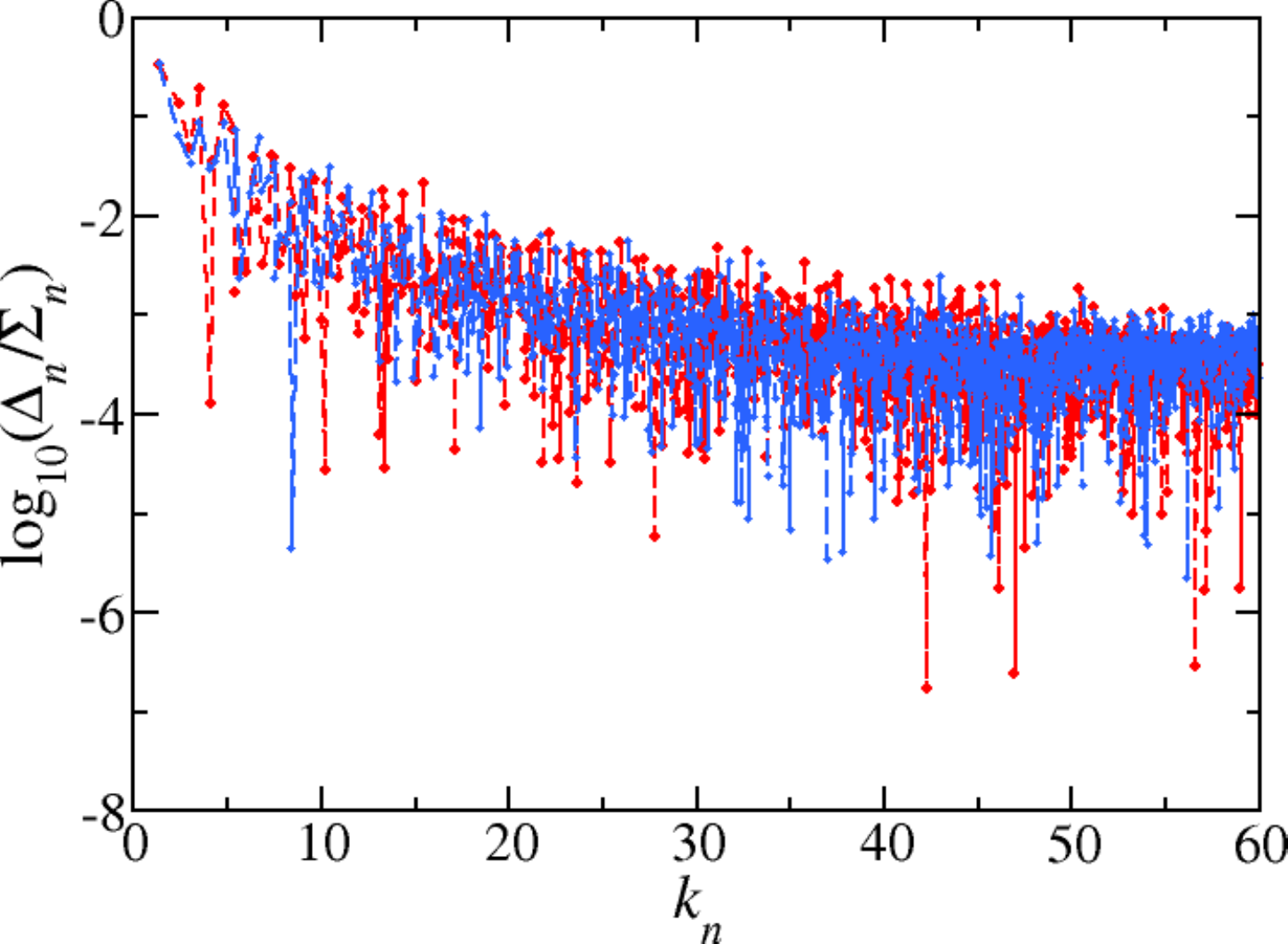}
\includegraphics[width=0.49\linewidth]{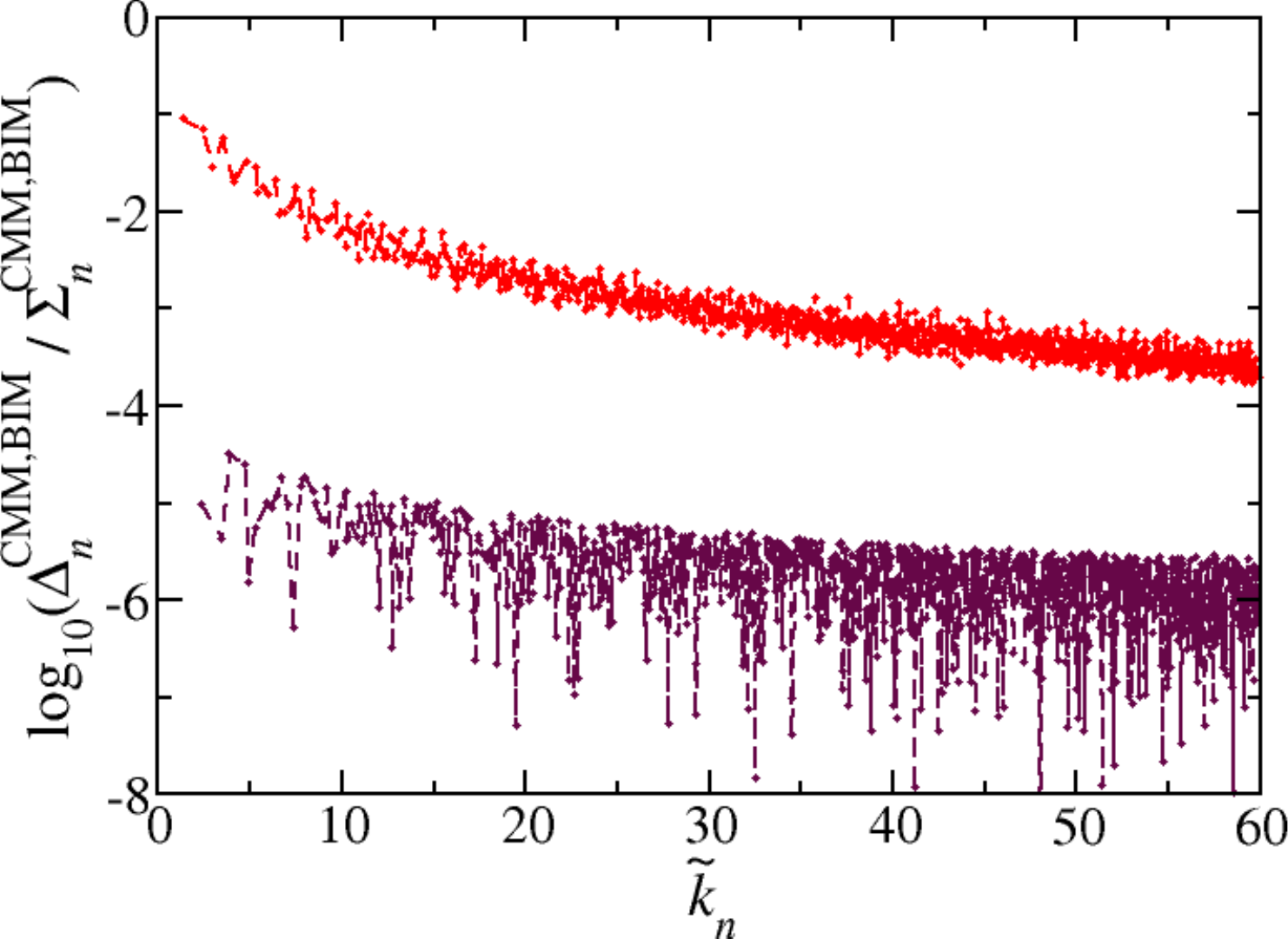}
	\caption{Left: Relative differences $\Delta_n/\Sigma_n=2\frac{|k^{a}_n-k^{b}_n|}{k^{a}_n+k^{b}_n}$ between the eigenvalues $k^a_n$ obtained from~\refeq{Schr2psi1} and $k^b_n$ from~\refeq{Schr2psi2} for the billiard with real $c_3$ (blue, left inset of~\reffig{Fig02}) and complex $c_3$ (red, right inset of~\reffig{Fig02}). Right: same as left for the eigenvalues $k^a_n$ of the shape~\refeq{WZ} with complex $c_3$ computed with the BIM and $k^b_n$ obtained with Eq.~(\ref{Solaa}) (red), and for the corresponding QB with eigenvalues $k^a_n=\tilde k_n$ computed with the BIM and $k^b_n$ obtained from~\refeq{CMMQB1} (maroon).
}
\label{Fig01}
\end{figure}

	It can be seen in~\reffig{Fig01} that the solutions of Eqs.~(\ref{Schr2psi1}) and~(\ref{Schr2psi2}) clearly deviate from each other for the shapes with real (blue) and complex (red) $c_3$, thus demonstrating that the assumption that the eigenvalues of the NB can be obtained from the sum of Eqs.~(\ref{Schr2psi1}) and~(\ref{Schr2psi2}) with $\boldsymbol{a}=\boldsymbol{b}$ is wrong. Also the solutions of~\refeq{Solaa} deviate from those obtained with the BIM for the corresponding NB, as exhibited in the right part of~\reffig{Fig02} (red). The high accuracy of the eigenvalues computed with the BIM can be deduced from the fact, that $\mathcal{\tilde A}=\mathcal{A}+o(10^{-6})$ and $\mathcal{\tilde L}\sim 10^{-6}$. Furthermore, to get an estimate for the size of deviations we also show the differences between the eigenvalues $k^a_n=\tilde k_n$ of the QB with complex $c_3$, computed with the BIM and the CMM~\refeq{CMMQB1} (maroon). These are a factor $10^{-3}$ smaller than for the NB. The spectral properties are close to GUE for the eigenvalues of the billiard with complex $c_3$, whereas they are close to the GOE for the shape with real $c_3$. This is expected, because for that case the shape has a mirror symmetry. However, in contrast to the non-relativistic QB, the associated spinor eigenstates cannot be separated according to their mirror symmetry, so the BIM and also the CMM should be applied to the whole billiard system. Yet, the spectral properties of the eigenvalues obtained from Eqs.~(\ref{Solaa}) and the BIM clearly deviate from each other, when considering for the former more than about 900 eigenvalues. This is illustrated in the right part of~\reffig{Fig02} for the shapes with real (blue) and complex (black) $c_3$ and for the Africa shape with $\alpha=0.2$. Shown are results for the number variance $\Sigma^2(L)$ in an interval of length $L$~\cite{Mehta2004} for the NB (open symbols) and the eigenvalues of~\refeq{Solaa}. As outlined above, these were unfolded with the Weyl formula for QBs (filled symbols), where the number of eigenvalues was chosen as large as possible, that is, such that the fluctuating part of the integrated spectral density fluctuates about zero, implying that there are no missing levels according to the unfolding procedure (open symbols). 
\section{Conclusions\label{Concl}}
In ~\cite{Xu2013} and subsequent works of these authors, the CMM is applied to NBs whose shapes originate from a conformal mapping of a circular shape. Starting point is an ansatz for the spinor components in terms of those of the eigenfunctions of the associated circular NB. These fulfill by construction the BC of that circular NB, but not that of the NB itself. Here, the expansion coefficients in the ansatz for the first spinor component are set equal to those of the second one, thereby ignoring their relation through the associated Dirac equation, and the Schr\"odinger equation~\refeq{Schr} is solved for their sum,~\refeq{Solaa}. We demonstrate in this work -- in the Secs.~\ref{Numerics} and~\ref{Example2} also numerically -- that equations~(\ref{Schr2psi1}),~(\ref{Schr2psi2}) for the spinor components and~\refeq{Solaa} do not have common solutions. Accordingly, the solutions of~\refeq{Solaa} are none of the Dirac equation for NBs and it is questionable to what extent they are connected to those of relativistic QBs. This discrepancy originates from the fact, that the conditional equation deduced from the BC~\refeq{BC2} for the circular NB differs from that for non-circular ones. Thus, the assumption, that the matrices entering the generalized eigenvalue problem~\refeq{Schr2psi1} are independent of $k$, is not applicable. On the contrary, for non-relativistic QBs the BC is independent of $\varphi$ and fulfilled for each term in the CMM expansion of the eigenstates in terms of those of the nonrelativistic circular QB [cf.~\refeq{ansatzpsi}]. 
\section{Acknowledgement}
The author acknowledges financial support from the Institute for Basic Science in Korea through the project IBS-R024-D1.
\bibliography{References_Comments}
\begin{appendix}
\begin{widetext}
\renewcommand{\theequation}{A\arabic{equation}}
\renewcommand{\thefigure}{A\arabic{figure}}
\setcounter{figure}{0}

\section{Appendix\label{suppl}}
\subsection{Some technical details} 
The plane-wave expansion for the second component Eq.~(14) in the main text is obtained by inserting the ansatz for the first component, Eq.~(13) into Eq.~(9) and employing the equalities 
\be
\label{Prop1}
\frac{1}{z^\ast[w^\prime (z)]^\ast}\left( r\frac{\partial}{\partial r} +i\frac{\partial}{\partial \varphi}\right)\vert w(z)\vert=e^{i\theta(z)},
\ee
\ba
\label{Prop2}
\frac{1}{z^\ast[w^\prime (z)]^\ast}\left( r\frac{\partial}{\partial r} +i\frac{\partial}{\partial \varphi}\right)e^{il\theta(z)}&=&-\frac{l}{\vert w(z)\vert} e^{i(l+1)\theta(z)},\\
\nonumber
\frac{1}{zw^\prime (z)}\left( r\frac{\partial}{\partial r} -i\frac{\partial}{\partial \varphi}\right)e^{il\theta(z)}&=&\frac{l}{\vert w(z)\vert} e^{i(l-1)\theta(z)},
\ea
and
\be
\label{Bessel}
J_{l-1}(x)=\frac{l}{x}J_l(x)+\frac{dJ_l(x)}{dx},\, J_{l+1}(x)=\frac{l}{x}J_l(x)-\frac{dJ_l(x)}{dx}.
\ee

The spinor eigenfunction,
\be
\boldsymbol{\Phi}_{n,\nu}(r,\varphi)=\begin{pmatrix}
	{\Phi_1}_{n,\nu}(r,\varphi) \\ {\Phi_2}_{n,\nu}(r,\varphi)
\end{pmatrix},
\ee
are orthogonal to each other, 
\ba
&&\int_0^{r_0}rdr\int_0^{2\pi}d\varphi\boldsymbol{\Phi}_{m,\mu}(r,\varphi)\cdot\boldsymbol{\Phi}_{n,\nu}(r,\varphi)\\
&&=\delta_{n,m}2\pi\int_0^{r_0}rdr\left[J_{m,\mu}(\kappa_{m,\mu}r)J_{m,\nu}(\kappa_{m,\nu}r)+J_{m+1,\mu}(\kappa_{m,\mu}r)J_{m+1,\nu}(\kappa_{m,\nu}r)\right]\mathcal{N}_{m,\mu}\mathcal{N}_{m,\nu}\\
&&\propto\delta_{n,m}\delta_{\nu,\mu}. 
\ea	
Namely, for ${\Phi_1}_{n,\nu}(r,\varphi)$ we have
\begin{align}
&\int_0^{r_0}rdr\int_0^{2\pi}d\varphi J_n(\kappa_{n,\nu} r)J_m(\kappa_{m,\mu} r)e^{i(n-m)\varphi},\\
&=2\pi\delta_{n,m}\frac{\kappa_{m,\nu}r_0J_{m+1}(\kappa_{m,\nu}r_0)J_m(\kappa_{m,\mu}r_0)-\kappa_{m,\mu}r_0J_{m+1}(\kappa_{m,\mu}r_0)J_m(\kappa_{m,\nu}r_0)}{\kappa^2_{m,\nu}-\kappa^2_{m,\mu}},\nonumber\\
&=2\pi\delta_{n,m}\left\{
\begin{array}{cll}
        \frac{J_m(\kappa_{m,\nu}r_0)J_m(\kappa_{m,\mu}r_0)}{\kappa_{m,\nu}+\kappa_{m,\mu}}&,&\nu\ne\mu\\
	-r_0^2J_{m}(\kappa_{m,\mu}r_0)J^\prime_{m,\mu}(\kappa_{m,\mu}r_0)&,&\nu=\mu\\
\end{array}\right.,\nonumber
\end{align}
and similarly for $\Phi_2(r,\varphi)$
\begin{align}
&\int_0^{r_0}rdr\int_0^{2\pi}d\varphi J_{n+1}(\kappa_{n,\nu} r)J_{m+1}(\kappa_{m,\mu} r)e^{i(n-m)\varphi},\\
&=2\pi\delta_{n,m}\frac{\kappa_{m,\mu}r_0J_{m}(\kappa_{m,\mu}r_0)J_{m+1}(\kappa_{m,\nu}r_0)-\kappa_{m,\nu}r_0J_{m}(\kappa_{m,\nu}r_0)J_{m+1}(\kappa_{m,\mu}r_0)}{\kappa^2_{m,\nu}-\kappa^2_{m,\mu}},\nonumber\\
&=2\pi\delta_{n,m}\left\{
\begin{array}{cll}
\frac{-J_m(\kappa_{m,\nu}r_0)J_m(\kappa_{m,\mu}r_0)}{\kappa_{m,\nu}+\kappa_{m,\mu}}&,&\nu\ne\mu\\
	r_0^2J_{m}(\kappa_{m,\mu}r_0)J^\prime_{(m+1),\mu}(\kappa_{m,\mu}r_0)&,&\nu=\mu\\
\end{array}\right.\, ,\nonumber
\end{align}
where we employed the BC Eq.~(20) in the main text, implying that for $\nu\ne\mu$ the sum of the contributions from the two spinor components cancel each other and
\be
\frac{1}{\mathcal{N}_{m,\mu}\mathcal{N}_{n,\nu}}\int_0^{r_0}rdr\int_0^{2\pi}d\varphi\boldsymbol{\Phi}_{m,\mu}(r,\varphi)\cdot\boldsymbol{\Phi}_{n,\nu}(r,\varphi)=\delta_{n,m}\delta_{\nu,\mu}\pi r_0^2\left.\left[\frac{dJ^2_{(m+1),\mu}(x)}{dx}-\frac{dJ^2_{m,\mu}(x)}{dx}\right]\right\vert_{(x=\kappa_{m,\mu})}=\frac{1}{\mathcal{N}^2_{n,\nu}}.
\ee
In analogy to the nonrelativistic case, Eq.~(19) in the main text, the matrices $\mathcal{\hat K}_1$, $\mathcal{\hat K}_2$, $\mathcal{\hat J}_1$ and $\mathcal{\hat J}_2$, are defined as
\ba
\label{MatrixDef}
&&\mathcal{\hat K}_{1,jj^\prime}=2\pi\delta_{nm}\mathcal{N}_{n,\nu}\mathcal{N}_{m,\mu}\int_0^{r_0}rdrJ_m(\kappa_{m,\mu}r)J_m(\kappa_{m,\nu}r)\kappa^2_{m,\nu},\\
&&\mathcal{\hat K}_{2,jj^\prime}=2\pi\delta_{nm}\mathcal{N}_{n,\nu}\mathcal{N}_{m,\mu}\int_0^{r_0}rdrJ_{m+1}(\kappa_{m,\mu}r)J_{m+1}(\kappa_{m,\nu}r)\kappa^2_{m,\nu},\nonumber\\
&&\mathcal{\hat J}_i{1,jj^\prime}=\mathcal{N}_{n,\nu}\mathcal{N}_{m,\mu}\int_0^{r_0}rdr\int_0^{2\pi}d\varphi\vert w^\prime(z)\vert^2J_m(\kappa_{m,\mu}r)J_n(\kappa_{n,\nu}r)i^{(n-m)}e^{i(n-m)\varphi},\nonumber\\
&&\mathcal{\hat J}_{2,jj^\prime}=\mathcal{N}_{n,\nu}\mathcal{N}_{m,\mu}\int_0^{r_0}rdr\int_0^{2\pi}d\varphi\vert w^\prime(z)\vert^2J_{m+1}(\kappa_{m,\mu}r)J_{n+1}(\kappa_{n,\nu}r)i^{(n-m)}e^{i(n-m)\varphi},\nonumber
\ea
with $[m,\mu]=[l(j),\lambda(j)],\, [n,\nu]=[l(j^\prime),\lambda(j^\prime)]$.
\subsection{Comparison of wave functions obtained from CMM and BIM for the Africa billiard with shape given in Eq. (E1) with $\alpha=0.2$}
\begin{figure}[!h]
\centering
\includegraphics[width=0.49\linewidth]{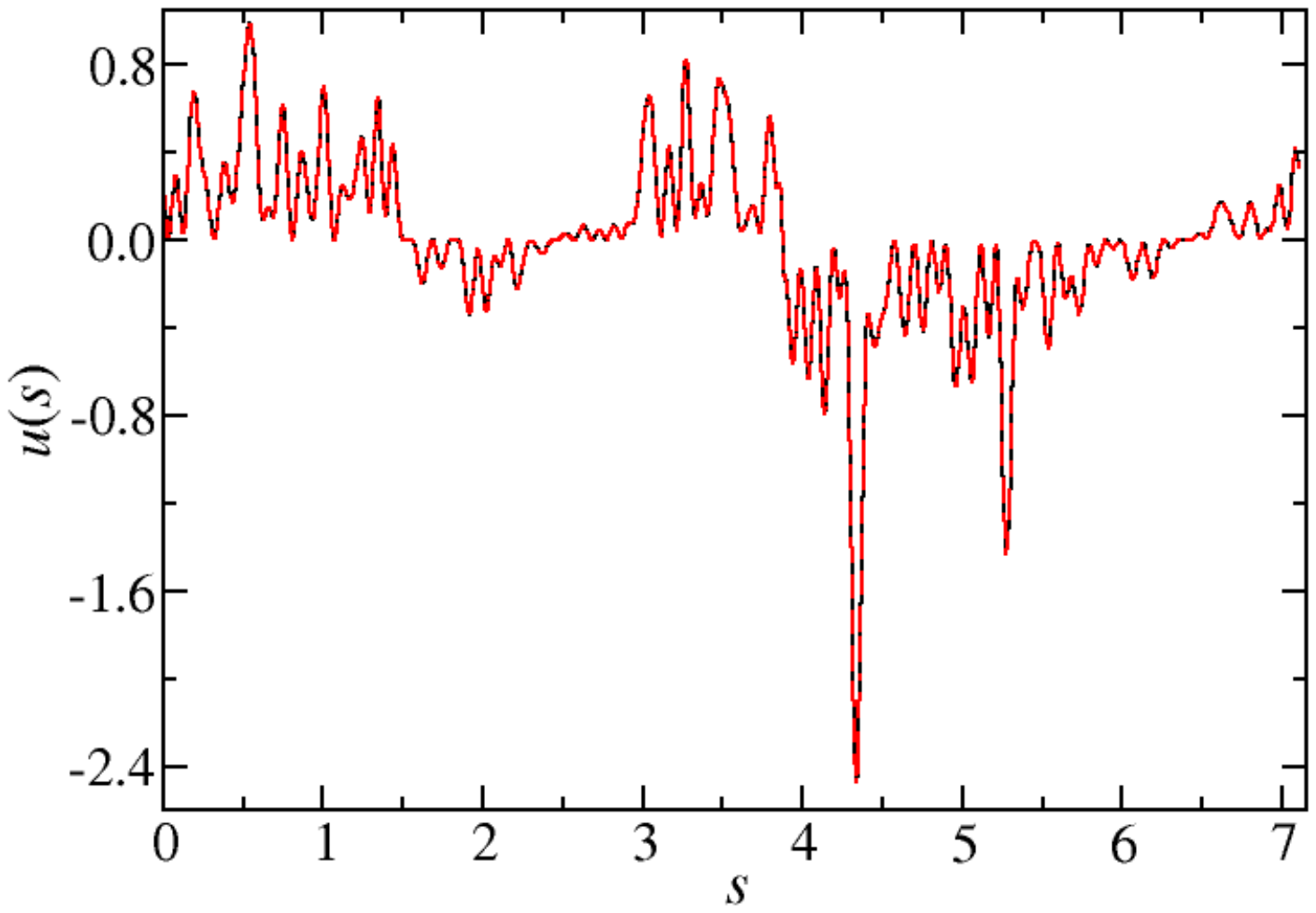}
\includegraphics[width=0.49\linewidth]{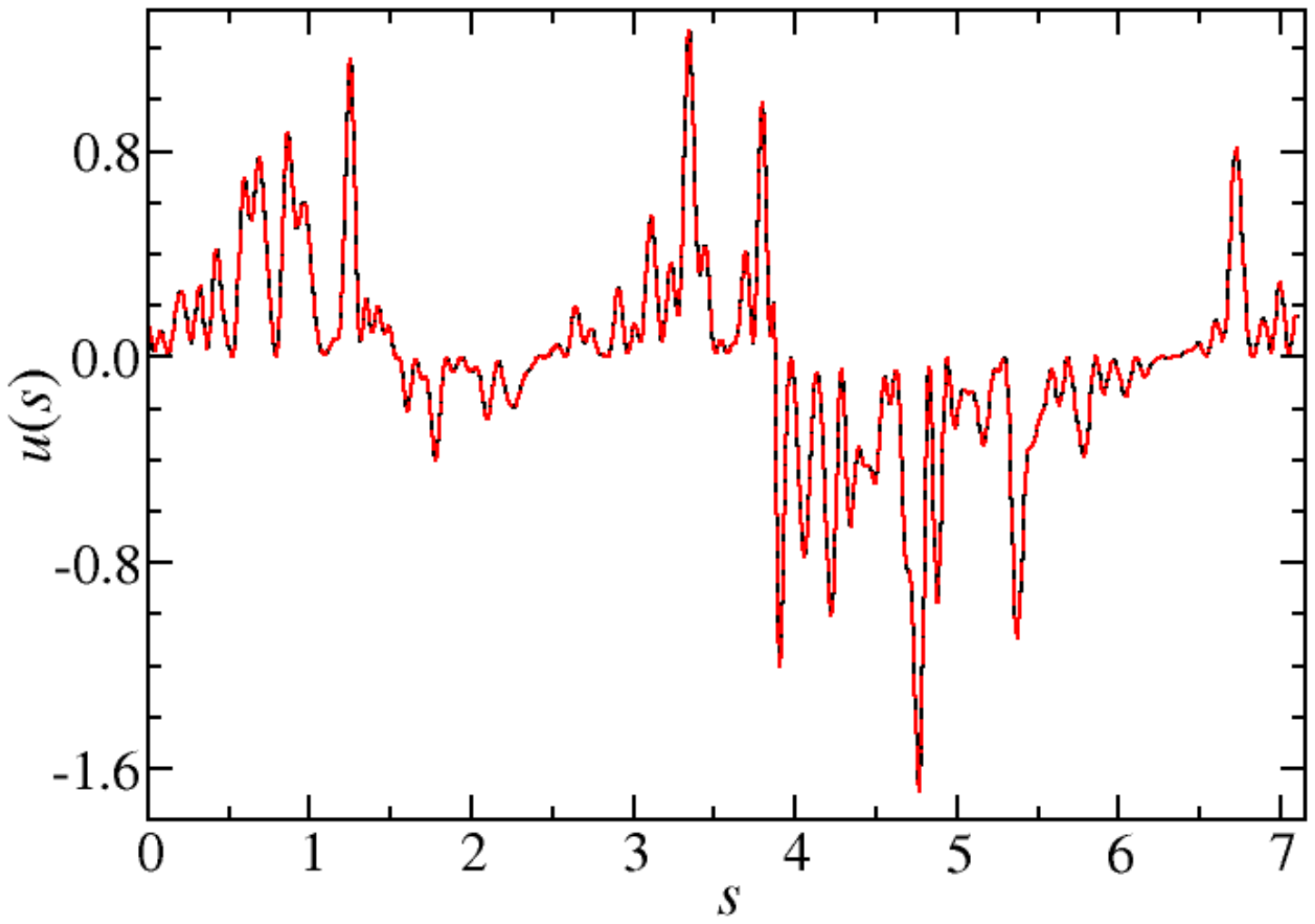}
	\caption{Outgoing flux along the boundary for the eigenstates 2507 (left) and 2508 (right) computed with CMM (black). Agreement with the analytical result Eq. (34) (red) of the main text is excellent.}
\label{Fig3}
\end{figure}
\begin{figure}[!h]
\centering
\includegraphics[width=0.23\linewidth]{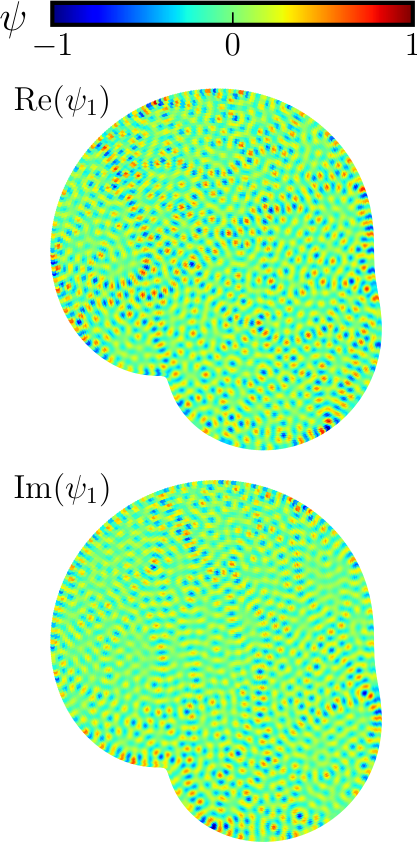}
\includegraphics[width=0.23\linewidth]{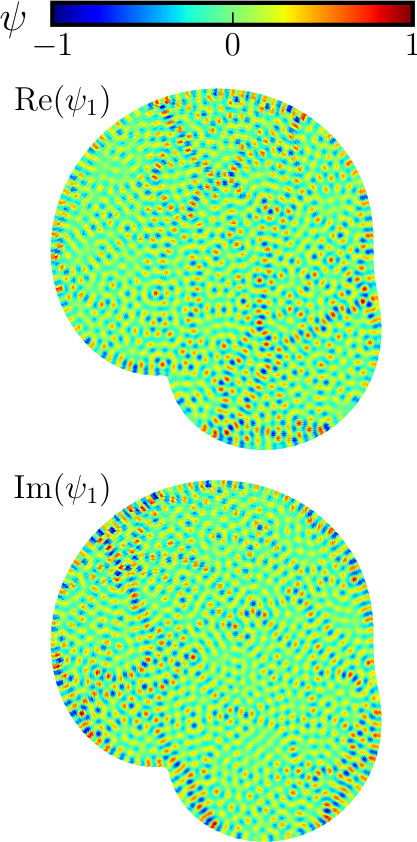}
\includegraphics[width=0.23\linewidth]{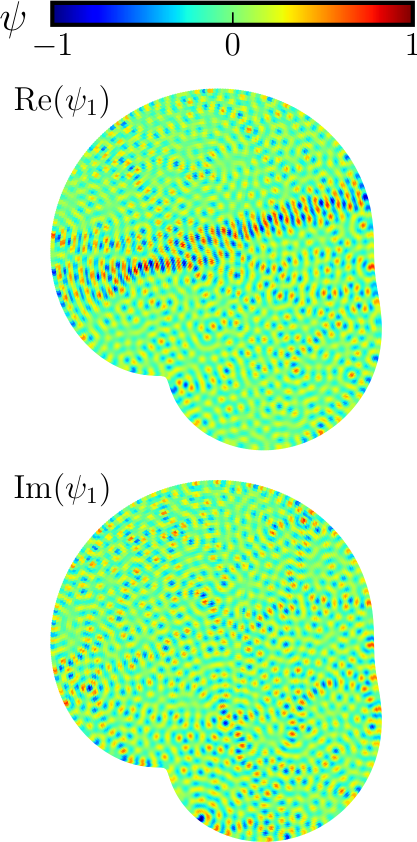}
\includegraphics[width=0.23\linewidth]{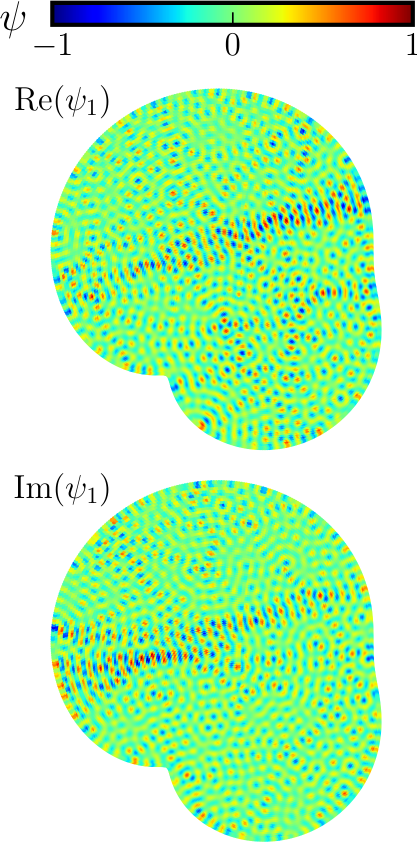}
\caption{Real and Imaginary parts of the first spinor component obtained from the BIM (1st and 3rd column) and CMM (2nd and 4th column) for the eigenstate 2507 (left pair of columns) and 2508 (right pair of column).}
\label{Fig4}
\end{figure}
\end{widetext}
\end{appendix}
\end{document}